\documentclass{Interspeech}

\usepackage{cite}
\usepackage{amsmath,amssymb,amsfonts}
\usepackage{algorithmic}
\usepackage{graphicx}
\usepackage{booktabs}
\usepackage{textcomp}
\usepackage{xcolor}
\usepackage{hyperref}

\usepackage{multirow}
\usepackage{graphicx,adjustbox}
\usepackage{siunitx}
\usepackage{caption}
\usepackage{tikz}
\usepackage{subcaption}

\interspeechcameraready 

\title{Accent Normalization Using Self-Supervised Discrete Tokens with Non-Parallel Data\thanks{*: Corresponding author.}}

\author[affiliation={1,2,4}]{Qibing}{Bai}
\author[affiliation={1,2}]{Sho}{Inoue}
\author[affiliation={3,1,2}]{Shuai}{Wang$^*$}
\author[affiliation={4}]{Zhongjie}{Jiang}
\author[affiliation={4}]{Yannan}{Wang}
\author[affiliation={1,2,5}]{Haizhou}{Li}

\newcommand{\combinedaffiliation}[4]{%
    \stepcounter{affcounter}%
    \edef\firstaffnum{\theaffcounter}%

    \stepcounter{affcounter}%
    \edef\secondaffnum{\theaffcounter}%

    \edef\affiliationslist{\affiliationslist\affiliationsep$^{\firstaffnum}$#1, $^{\secondaffnum}$#2, #3, #4}%
    \edef\metadataaffiliations{\metadataaffiliations \{\firstaffnum\}\{#1\}\{#3\}\{#4\}; }%
    \edef\metadataaffiliations{\metadataaffiliations \{\secondaffnum\}\{#2\}\{#3\}\{#4\}; }%
    \renewcommand{\affiliationsep}{\vskip1pt}%
}

\combinedaffiliation{School of Data Science}{SRIBD}{The Chinese University of Hong Kong, Shenzhen}{China}
\affiliation{School of Intelligence Science and Technology}{Nanjing University, Suzhou}{China}
\affiliation{Tencent Ethereal Audio Lab}{Tencent}{China}
\affiliation{Department of ECE}{National University of Singapore}{Singapore}

\email{}


\usepackage{comment}

\begin{document}

\maketitle

\begin{abstract}
Accent normalization converts foreign-accented speech into native-like speech while preserving speaker identity. We propose a novel pipeline using self-supervised discrete tokens and non-parallel training data. The system extracts tokens from source speech, converts them through a dedicated model, and synthesizes the output using flow matching. Our method demonstrates superior performance over a frame-to-frame baseline in naturalness, accentedness reduction, and timbre preservation across multiple English accents. Through token-level phonetic analysis, we validate the effectiveness of our token-based approach. We also develop two duration preservation methods, suitable for applications such as dubbing.
\footnote{Samples: {\url{https://p1ping.github.io/TokAN-Demo}}. Code available at {\url{https://github.com/P1ping/TokAN}}.}
\end{abstract}

\section{Introduction}
Accent conversion (AC) seeks to alter speech from one accent to another while preserving the speaker characteristics. We study the way to convert the non-native (L2) accented speech into a native (L1) accented one. Such accent normalization (AN)\footnote{Also referred to as foreign accent conversion (FAC).} technology offers a wide range of practical applications, such as improving pronunciation for language learners~\cite{felps2009foreign}, enhancing the authenticity of movie dubbing \cite{turk2002subband}, and personalized text-to-speech (TTS) systems \cite{sun2016personalized}.

Early deep learning approaches focused on \textit{reference-based methods}~\cite{zhao2018icassp, zhao2019foreign, li2020improving, ding2022accentron}, which rely on native accent speech samples to generate accent-neutral representations. Some studies~\cite{zhao2018icassp, zhao2019foreign, ding2022accentron} use pre-trained automatic speech recognition (ASR) models to extract phonetic posteriorgram (PPG) or bottleneck features, while \cite{li2020improving} propose to generate reference speech with native text-to-speech (TTS). A recent method~\cite{zhang2025vevo} leverages speech tokens for style transfer. However, the requirement for reference speech during inference limits their practical applications.

The \textit{reference-free method}~\cite{zhao2021converting, nguyen2022accent, quamer2022zero} has gained attention, focusing on the mapping between non-native and native accent representations~\cite{zhao2021converting, nguyen2022accent} or converting PPGs for unseen speakers~\cite{quamer2022zero}. \cite{huang2023evaluating} provides a comprehensive evaluation, while \cite{jia2024convert} proposes to use discrete self-supervised tokens with small-scale paired data. Although these approaches eliminate reference speech requirements, they still rely on paired parallel data during training. The challenges in assembling diverse accent corpora have inspired the development of methods that are both reference-free and independent of parallel corpora~\cite{liu2020end, jin2023voice, zhou2023tts, jia2022zero, chen2024transfer}.

Recent accent normalization approaches address parallel data limitations. \cite{liu2020end} proposes a two-step process: converting non-native speech to text via ASR, and then generating native speech using TTS. \cite{jin2023voice} introduces a pseudo siamese network with an auxiliary decoder to separate accent features. \cite{zhou2023tts} leverages TTS representations as accent-neutral features. Building on this, \cite{chen2024transfer} incorporates a non-autoregressive (NAR) architecture and enhances robustness through pre-training techniques. The TTS-guided framework was then augmented with flow matching and normalizing flow~\cite{bai2024diffusion,nguyen2024improving}, respectively.

Recent work \cite{zhou2023tts,chen2024transfer,bai2024diffusion,nguyen2024improving} relies on the TTS-synthesized targets, and the quality of synthetic targets could be a limitation, both in terms of voice cloning and duration modeling. When the source speech is noisy, it is hard to generate target speech with the same voice. Moreover, the duration is learned from the synthetic targets instead of native utterances, this injects error accumulation and also applies to PPG-based methods.

We propose an accent normalization method using self-supervised discrete tokens and non-parallel data. Our method first quantizes both input speech and TTS-generated targets into discrete tokens, which reduces quality requirements for synthetic targets. A token-to-token conversion model, trained on deduplicated tokens where consecutive identical tokens are merged, transforms source-accented sequences into native ones. This deduplication removes duration information and focuses the model on the phonetic mapping between accents. A flow matching token-to-Mel synthesizer then recovers duration and generates Mel-spectrograms.
Additionally, we explore two methods for preserving the utterance duration: a basic scaling approach and a flow-matching-based technique, which are particularly valuable for applications such as dubbing.
Our contributions are three-fold:
\begin{itemize}
\item We perform accent normalization using self-supervised discrete tokens and non-parallel data, mitigating the impact of synthetic data quality through token-only training.
\item We validate our approach across multiple English accents through comprehensive evaluations, supported by a token-level phonetic analysis.
\item We explore duration modeling techniques that preserve the total duration of the original utterance
\end{itemize}

\section{Related Work}

\begin{figure*}[t]
\centering
\includegraphics[width=0.95\textwidth]{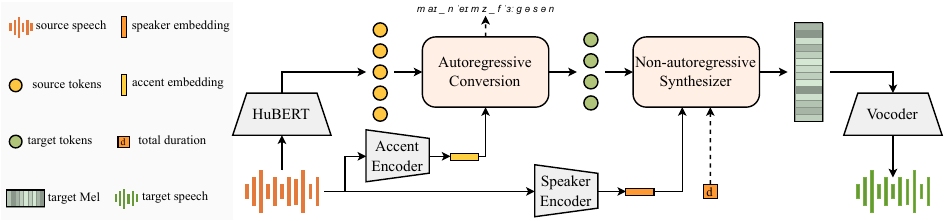}
\caption{Pipeline of accent normalization. Trainable modules include the conversion model and synthesizer, while gray trapezoids indicate pre-trained modules. The conversion model can generate auxiliary phone outputs. The synthesizer, with a flow matching duration predictor, enables total duration control.}
\vspace{-0.5cm}
\label{fig:pipeline}
\end{figure*}

\subsection{Self-Supervised Discrete Tokens in Speech Generation}
Discrete tokens from self-supervised learning (SSL)~\cite{hsu2021hubert} exhibit a strong correlation with phonetic content~\cite{choi24self}, making them effective representations for speech synthesis. These tokens serve as a proxy in various generation tasks, including text-to-speech (TTS)~\cite{kharitonov2023speak} and voice conversion (VC)~\cite{huang2021any}. Their discrete nature enables the application of textual techniques to speech processing, such as spoken language modeling~\cite{lakhotia2021generative}, speech-to-speech translation~\cite{lee2022direct}, and speech generation in large language models~\cite{fang2024llamaomni}. We leverage HuBERT~\cite{hsu2021hubert} to extract discrete tokens for conversion and synthesis.

\subsection{Flow-Matching-Based Speech Synthesis}
Flow matching~\cite{lipman2023flow} has emerged as a powerful approach for acoustic feature generation in TTS systems~\cite{mehta2024matcha}, achieving competitive naturalness. Recent non-autoregressive TTS research explores flow matching duration predictors~\cite{le2023voicebox,ju2024naturalspeech,eskimez24total}. While these predictors show higher word error rates (WER) compared to regression, they enable more diverse rhythmic patterns~\cite{le2023voicebox}. We apply flow matching to accent normalization.
While using regression-based duration prediction by default, we investigate a flow matching duration predictor with awareness of the total duration.

\section{Methodology}

Figure~\ref{fig:pipeline} illustrates our accent normalization pipeline. The system first converts input speech into discrete HuBERT tokens through feature extraction, quantization, and deduplication. An autoregressive conversion model then transforms these source tokens into target tokens with normalized phonetic patterns, removing the source accent. A non-autoregressive synthesizer predicts token-level durations and generates the Mel-spectrogram, conditioned on the source speaker embedding. The synthesizer can employ a flow matching duration predictor for total duration control.

We train the conversion model and synthesizer on separate datasets. The synthesizer is trained on native-accented speech to model normalized duration and acoustic features, while the conversion model requires parallel data to remove source accents. Since token representations largely exclude timbre information, content-parallel training data suffices. This token-based approach prevents timbre or prosodic artifacts in synthetic training speech from degrading target token quality.

\subsection{Autoregressive Token Conversion}

HuBERT's discrete tokens primarily encode content-related information, with minimal paralinguistic features when extracted from appropriate layers~\cite{hsu2021hubert}. These unsupervised tokens capture phonetic rather than semantic patterns~\cite{choi24self}, preserving source accent characteristics in the token sequence. We employ an autoregressive model to normalize these accented patterns.

While the conversion model typically requires text-parallel data, such pairs are challenging to collect. For non-parallel L2-accented datasets, we generate synthetic targets using a native TTS.
Previous approaches~\cite{chen2024transfer,bai2024diffusion,nguyen2024improving} using synthetic data rely on direct speech supervision, which can lead to error accumulation during training. Our approach, operating on deduplicated tokens, mitigates this issue by eliminating most paralinguistic and duration information from synthetic data.

To enhance content understanding, we incorporate CTC-based~\cite{graves2006connectionist} phone prediction as an auxiliary training task, complementing the model's implicit content recognition from input tokens. The model's handling of diverse input accents is further improved through accent-conditional processing, where accent embeddings from a pre-trained classifier are integrated into the encoder using AdaLN~\cite{peebles2023scalable}.

\subsection{Non-Autoregressive Token-to-Mel Synthesis}\label{sec:token_to_mel}

The synthesizer maps native-normalized tokens to speech through a non-autoregressive architecture comprising a duration predictor, token encoder, and Mel decoder. Using flow matching~\cite{li2020improving}, the system processes input tokens and speaker embedding to generate speech in multiple steps.

First, the Transformer-based token encoder converts discrete tokens into continuous representations, conditioned on speaker embeddings. The duration predictor then assigns duration values to these token representations for expansion. The expanded representations serve as the condition for the Mel decoder, which transforms noisy Mel-spectrograms into high-likelihood ones via velocity prediction. The decoder employs a UNet-Transformer architecture~\cite{mehta2024matcha} with AdaLN-Zero~\cite{peebles2023scalable} speaker conditioning, enhanced by classifier-free guidance (CFG)~\cite{ho2021classifierfree} for improved speech quality.

\textbf{Preservation of total duration.}
While the default 1-D convolutional duration predictor, trained with mean square error, generates robust durations~\cite{le2023voicebox}, it offers limited diversity and control. For applications like dubbing where matching source duration is crucial, we explore two approaches. The first simply scales predicted durations uniformly to match the target total duration (denoted as ``ours w/ dur. scaling" in experiments).

We also develop a flow matching duration predictor conditioned on \textit{average token duration}, enabling direct total duration control. This approach generates durations that naturally sum closer to the target length, requiring minimal scaling adjustment. Our experiments compare both methods, with this total-duration-aware system labeled as ``ours w/ dur. control".

\subsection{Training Pipeline}

Since discrete HuBERT tokens serve as the intermediary between the conversion model and the synthesis model, the two models can be trained independently using distinct datasets. The synthesizer is directly trained on high-quality target-accented data, typically a TTS dataset.
The conversion model is trained on non-parallel data with synthetic targets:
\begin{itemize}
    \item Train a TTS model on native-accented data
    \item Generate TTS-synthesized targets for L2-accented data
    \item Extract discrete SSL tokens for both the source and target
    \item Train the conversion model on these token pairs
\end{itemize}
\noindent
In line with \cite{kharitonov2023speak,jia2024convert}, BART-style pre-training~\cite{lewis2019bart} is employed to enhance model performance, additionally with the phone supervision.
This modular design enables the use of different datasets optimized for each component while maintaining the effectiveness of the overall system.

\section{Experimental Setup}

\subsection{Datasets}\label{sec:data}

We evaluate our method on seven English accents: six from L2-ARCTIC~\cite{zhao2018l2arctic} (Arabic, Chinese, Hindi, Korean, Spanish, and Vietnamese) and native American English from ARCTIC~\cite{kominek2004cmu}. Each accent consists of four speakers. We conduct evaluations on this extended L2-ARCTIC dataset.
For training, we utilize multiple data sources:
\begin{itemize}
    \item LibriTTS-R~\cite{koizumi2023librittsr} for the auxiliary TTS model and synthesizer
    \item The extended L2-ARCTIC described above ($\sim$20 hours)
    \item A proprietary Chinese-accented ASR dataset ($\sim$~300 hours)
\end{itemize}
LibriTTS-R is also used to pre-train the conversion model. The proprietary Chinese-accented dataset is used exclusively for training and validation. Since L2-ARCTIC uses the same prompt sentences across speakers, we construct the data splits: 50 sentences for validation and 80 for testing, holding out one unseen speaker per accent.

\subsection{Model Architecture}

\textbf{Proposed.} We employ the HuBERT-large model\footnote{\url{https://github.com/facebookresearch/fairseq/blob/main/examples/hubert}} (17th layer features) with K-means clustering on LibriTTS-R for 1000 discrete tokens. The conversion model follows an encoder-decoder architecture (each with 5 layers and a hidden dimension of 1024) with accent information from a pretrained classifier\footnote{\url{https://huggingface.co/Jzuluaga/accent-id-commonaccent_xlsr-en-english}}. For synthesis, we use flow matching~\cite{lipman2023flow} with Resemblyzer\footnote{\url{https://github.com/resemble-ai/Resemblyzer}} speaker embeddings and BigVGAN~\cite{lee2023bigvgan} vocoding.

\noindent
\textbf{Baseline.} We implement a frame-to-frame conversion approach following~\cite{bai2024diffusion}. The model consists of three training stages: TTS training, speech encoder pre-training via distillation, and speech encoder fine-tuning. We utilize pitch and energy prediction during inference for more native prosody.

\subsection{Training \& Inference Configurations}

For synthetic target generation, we employ Matcha-TTS~\cite{mehta2024matcha} with voice cloning for the extended L2-ARCTIC dataset and a fixed speaker embedding for the extra Chinese-accented data.

The conversion model pre-training uses token noise with probabilities of 0.2 for masking, 0.1 for insertion, and 0.1 for replacement. We set the CTC loss weight to 1.0. The synthesizer training applies a CFG rate of 0.2 for both decoder and duration predictor.

During inference, we use a beam size of 10 for the conversion model. The token-to-Mel synthesizer operates with 32 sampling steps and CFG strengths of 2.0 and 0.5 for the Mel decoder and flow matching duration predictor, respectively.

\subsection{Evaluation Metrics}\label{sec:metrics}

\textbf{Subjective Metrics.} We evaluate speech naturalness (NAT) and accentedness (ACT) using MUSHRA tests, and speaker similarity (SIM) using best-worst scaling (BWS). The BWS results are aggregated via the counting algorithm~\cite{ravillion2020comparison}: $(N_\textit{best} - N_\textit{worst}) / N_\textit{occurrence}$. Each evaluation involves 21 raters assessing 20 samples, with the American accent excluded from the accentedness evaluation.

\noindent
\textbf{Objective Metrics.}
We assess three aspects of conversion quality. For \textit{intelligibility}, we compute the word error rate (WER) using a native-only ASR model\footnote{\url{https://huggingface.co/facebook/s2t-medium-librispeech-asr}}, simulating listener's perception of L2-accented speech. \textit{Timbre preservation} is measured by speaker encoding similarity (SECS) using Resemblyzer, which exhibits relatively high accent-independence compared to other speaker models.
For \textit{accentedness reduction}, we compute metrics against the synthetic targets. Following the three aspects of accent (pronunciation, rhythm, and intonation)~\cite{zhou2024accented}, we employ PPG distance ($\Delta\text{PPG}$)~\cite{churchwell2024high}, frame distortion (FD), and F0 correlation (F0 corr.), respectively.

\begin{table*}[htbp]
  \centering
  \caption{Evaluation results of the accent conversion systems.}
  \vspace{-0.3cm}
  \setlength{\tabcolsep}{7pt}  
  \small  
  \resizebox{0.98\textwidth}{!}{  
  \begin{tabular}{lccccccccc}
    \toprule
    \multirow{2}{*}{System} & \multirow{2}{*}{Source-length} & \multicolumn{3}{c}{Subjective}&\multicolumn{5}{c}{Objective} \\ \cmidrule(r){3-5}\cmidrule(r){6-10}
    & & NAT ($\uparrow$) & ACT ($\downarrow$) & SIM ($\uparrow$) & WER (\% $\downarrow$) & SECS ($\uparrow$) & $\Delta\text{PPG}$ ($\downarrow$) & FD ($\downarrow$) & F0 corr. ($\uparrow$) \\
    \midrule
    Source & $\checkmark$ & 60.50{\tiny$\pm$2.34} & 46.40{\tiny$\pm$2.48} & - & 15.86 & - & 0.51 & 67.29 & 0.74 \\
    Source resynthesis & $\checkmark$ & - & - & - & 18.75 & 0.8863 & 0.47 & 66.79 & 0.75 \\
    \midrule 
    Baseline~\cite{bai2024diffusion} & $\checkmark$ & 48.76{\tiny$\pm$2.36} & 37.61{\tiny$\pm$2.07} & $-$0.031 & 21.54 & 0.7846 & 0.49 & 60.68 & 0.72 \\
    Ours & $\times$ & {\bf 60.38}{\tiny$\pm$2.17} & {\bf26.14}{\tiny$\pm$1.79} & $\phantom{-}${\bf0.040} & {\bf 16.25} & 0.8739 & {\bf 0.30} & {\bf 35.24} & {\bf 0.77} \\
    Ours w/ dur. scaling & $\checkmark$ & 54.24{\tiny$\pm$2.22} & 26.80{\tiny$\pm$1.79} & $\phantom{-}$0.007 & 16.47 & 0.8771 & 0.31 & 62.71 & 0.77 \\
    Ours w/ dur. control & $\checkmark$ & 54.72{\tiny$\pm$2.23} & 26.61{\tiny$\pm$1.82} & $-$0.017 & 16.76 & {\bf 0.8774} & 0.32 & 61.57 & 0.77 \\
    \bottomrule
  \end{tabular}
  }  
\vspace{-0.4cm}
\label{tab:res}
\end{table*}

\section{Results}

\subsection{Comparison with Baseline}

\textbf{Naturalness and speaker similarity.}
Table~\ref{tab:res} presents both subjective and objective evaluation results.
The proposed system outperforms the frame-to-frame baseline in naturalness and speaker similarity, both subjectively and objectively. This might be attributed to i) the error accumulation during the baseline training process, and ii) the use of CFG in the proposed synthesizer, which influences speaker similarity in particular. 

\noindent
\textbf{Accentedness and intelligibility.}
The proposed system achieves more effective accent normalization, particularly through its flexible adjustment of speech rhythm, a critical component of accent. Both systems exhibit higher WERs compared to the source speech, indicating challenges in maintaining intelligibility during accent conversion.
The native-only ASR model's high WERs for the baseline system reflect its lower intelligibility, likely due to its less effective accent reduction.

Table~\ref{tab:accent_wise_act} shows the accent-wise results for the subjective evaluation of accentedness. Both the baseline and proposed model can reduce the accentedness on every accent, while the proposed model consistently outperforms the baseline.

\begin{table}[htbp]
  \centering
  \caption{Accent-wise subjective results of accentedness.}
  \vspace{-0.3cm}
  \label{tab:accent_wise_act}
  \setlength{\tabcolsep}{1pt}  
  \small
  \resizebox{0.46\textwidth}{!}{
  \begin{tabular}{lcccccc}
    \toprule
    System & Zh & Hi & Vi & Ar & Es & Ko \\
    \midrule
    Source & 46.20{\tiny$\pm$5.25} & 59.65{\tiny$\pm$7.02} & 43.53{\tiny$\pm$5.66} & 52.61{\tiny$\pm$6.14} & 44.79{\tiny$\pm$6.04} & 33.54{\tiny$\pm$4.80} \\
    \midrule
    Baseline & 39.09{\tiny$\pm$4.78} & 41.82{\tiny$\pm$5.81} & 40.56{\tiny$\pm$5.59} & 34.94{\tiny$\pm$4.71} & 37.99{\tiny$\pm$5.52} & 32.76{\tiny$\pm$4.21} \\
    Ours & {\bf30.41}{\tiny$\pm$4.98} & {\bf28.03}{\tiny$\pm$4.70} & {\bf27.49}{\tiny$\pm$5.94} & {\bf23.54}{\tiny$\pm$4.46} & {\bf24.18}{\tiny$\pm$3.69} & {\bf24.34}{\tiny$\pm$3.30} \\
    \bottomrule
  \end{tabular}
  }
\vspace{-0.2cm}
\end{table}

The native-only ASR model's WER reflects the intelligibility of converted speech from a native-only listener's perspective. As shown in Table~\ref{tab:accent_wise_wer}, our model outperforms the baseline, particularly for Chinese and Vietnamese accents. This improvement may be attributed to our method's ability to handle syllable-timed rhythm patterns from these L1 languages.
For the Hindi accent, however, our method has lower performance, likely due to Hindi's fine-grained features (e.g., retroflex consonants) that are not well captured in our discrete token inventory.

\begin{table}[htbp]
  \centering
  \caption{Accent-wise WERs with native-only ASR}
  \vspace{-0.3cm}
  \label{tab:accent_wise_wer}
  \setlength{\tabcolsep}{5pt}  
  \small
  \resizebox{0.46\textwidth}{!}{
  \begin{tabular}{lccccccc}
    \toprule
    System & Zh & Hi & Vi & Ar & Es & Ko & Us \\
    \midrule
    Source & 20.82 & 11.75 & 31.40 & 15.09 & 15.79 & 13.78 & 2.53 \\
    \midrule
    Baseline & 26.76 & {\bf19.98} & 34.36 & 22.96 & 22.28 & 18.91 & 5.62 \\
    Ours & {\bf17.51} & 24.93 & {\bf18.73} & {\bf17.33} & {\bf17.05} & {\bf13.94} & {\bf5.55} \\
    \bottomrule
  \end{tabular}
  }
\vspace{-0.5cm}
\end{table}

\subsection{Token Distributions of Phonemes}

\begin{figure}[htbp]
\centering
\includegraphics[width=0.45\textwidth]{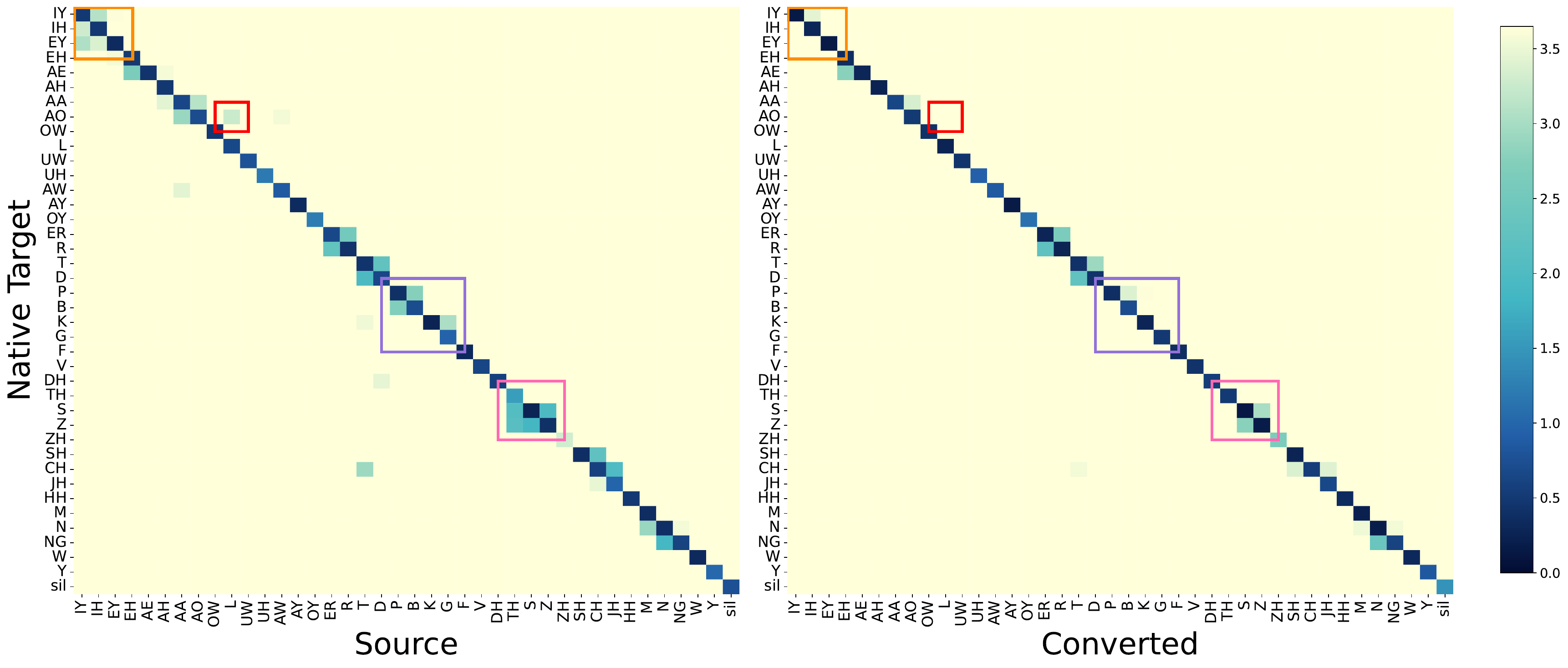}
\vspace{-0.2cm}
\caption{Token-based phoneme divergence from native targets. Left: source Chinese-accented speech. Right: converted speech.}
\vspace{-0.6cm}
\label{fig:distribution}
\end{figure}

To evaluate pronunciation normalization, we analyze phoneme-specific token distributions using forced alignment on the test set. Computing KL divergence between these distributions and those from American native speech reveals the effectiveness of our conversion approach. Figure~\ref{fig:distribution} illustrates the analysis for Chinese-accented speech before and after conversion.

The post-conversion results show a more prominent low-valued diagonal line, indicating successful transformation toward native-like pronunciations. Significant improvements appear in common Chinese-English mispronunciation patterns: high-front vowels (orange), substitution of \textit{AO} for dark \textit{L} (red), confusion between voiced and unvoiced plosives (purple), and dental-alveolar fricative distinctions (pink).

\subsection{Total-Duration Preservation}

\begin{figure}[htbp]
\centering
\includegraphics[width=0.40\textwidth]{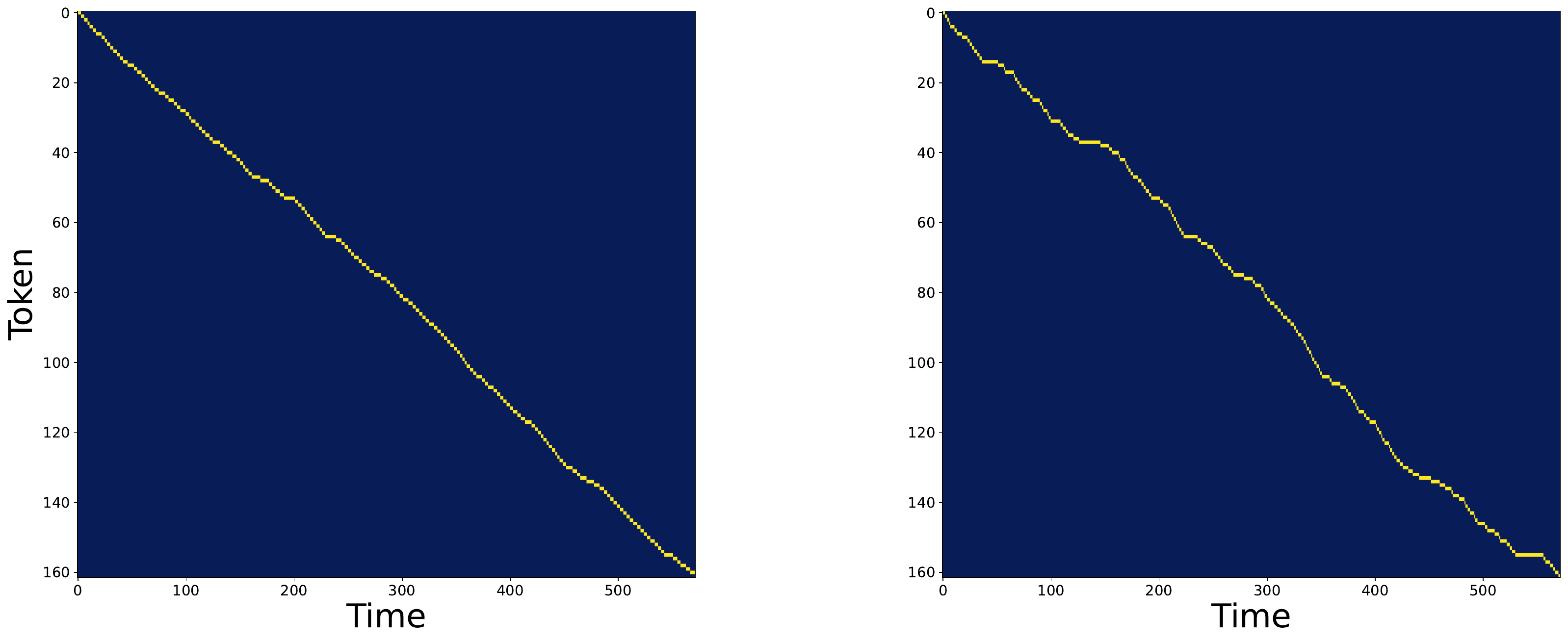}
\vspace{-0.3cm}
\caption{Comarison of two total-duration controlling methods: ``ours w/ dur. scaling" (left) and ``ours w/ dur. control" (right).}
\label{fig:alignment}
\vspace{-0.5cm}
\end{figure}

As detailed in Sec.~\ref{sec:token_to_mel}, we compare two duration control methods: direct scaling of predicted durations (``ours w/ dur. scaling") and a flow matching duration predictor with total-duration awareness (``ours w/ dur. control"). Without scaling, our base model (``ours") shows an average absolute difference of $0.513$ seconds (around $12.89\%$) from source durations on the testing set. In contrast, The duration control approach achieves an absolute difference of $0.064$ seconds (around $1.64\%$). Figure~\ref{fig:alignment} demonstrates their typical behavior. The scaling approach produces more uniform durations, shown by the diagonal alignment, while the control approach yields more varied durations, particularly for vowels and fricatives.

While ``ours w/ dur. control" achieves less frame disturbance and similar SECS scores compared to direct scaling, subjective SIM ratings favor the scaling approach. This preference likely stems from their different handling of source disfluencies: scaling preserves the slower speaking rate, while the control method produces a more acted speaking style that may affect perceived speaker identity.

\subsection{Ablation Study}

\noindent
\textbf{Auxiliary modules.} Removing accent embeddings or phone supervision leads to noticeable performance degradation, as shown by increased WER in Table~\ref{tab:ablation_res}. Phone supervision proves particularly crucial for content preservation.

\noindent
\textbf{Training data.} Removing the extra Chinese-accented data results in a slight performance decline.  Removing the pre-training stage causes significant performance degradation. This suggests that the pre-training step is crucial for the model to learn various linguistic patterns and native token distributions.

\begin{table}[thbp]
  \centering
  \caption{Ablation results for conversion model}
  \vspace{-0.3cm}
  \label{tab:ablation_res}
  \setlength{\tabcolsep}{4pt}  
  \small
  \resizebox{0.45\textwidth}{!}{
  \begin{tabular}{lcccc}
    \toprule
    System & WER ($\% \downarrow$) & $\Delta\text{PPG}$ ($\downarrow$) & FD ($\downarrow$) & F0 corr. ($\uparrow$)  \\
    \midrule
    Source & 15.86 & 0.51 & 67.29 & 0.74 \\
    \midrule
    Ours & {\bf 16.25} & {\bf 0.30} & 35.24 & {\bf 0.77} \\
    \ - accent embeddings & 16.82 & 0.31 & 35.75 & 0.76 \\
    \quad - phone supervision & 19.55 & 0.33 & 34.97 & 0.77 \\
    \ - extra data & 17.13 & 0.31 & 39.75 & 0.77 \\
    \ - pre-training & 27.59 & 0.40 & {\bf 30.83} & 0.76 \\
    \bottomrule
  \end{tabular}
  }
\vspace{-0.3cm}
\end{table}

\section{Conclusion, Limitation, \& Future Work}

In this paper, we introduce an accent normalization method using self-supervised discrete tokens and non-parallel data. Subjective evaluations on multiple English accents show significant improvements in speech naturalness, speaker similarity, and accentedness reduction. We also investigate two methods for controlling total duration, suitable for dubbing scenarios. Subjective evaluations indicate a preference for direct duration-scaling, whereas the flow matching method generates dynamic durations close to the total duration.

Despite improved naturalness and accent normalization, the high WER after conversion limits practical use. Future work will explore more robust tokenizers to address this limitation.


\section{Acknowledgements}
This work was supported by National Natural Science Foundation of China (Grant No. 62401377), Shenzhen Science and Technology Program (Shenzhen Key Laboratory, Grant No. ZDSYS20230626091302006), Shenzhen Science and Technology Research Fund (Fundamental Research Key Project, Grant No. JCYJ20220818103001002), Program for Guangdong Introducing Innovative and Entrepreneurial Teams (Grant No. 2023ZT10X044).

{\scriptsize
\bibliographystyle{IEEEtranEtAl}
\bibliography{mybib}
}

\end{document}